\documentclass{article}

\usepackage{PRIMEarxiv}

\usepackage[utf8]{inputenc} 
\usepackage[T1]{fontenc}    
\usepackage{hyperref}       
\usepackage{url}            
\usepackage{booktabs}       
\usepackage{amsfonts}       
\usepackage{nicefrac}       
\usepackage{microtype}      
\usepackage{lipsum}
\usepackage{fancyhdr}       
\usepackage{graphicx}       
\graphicspath{{media/}}  
\usepackage{amsmath}

\def\##1{\underline #1}
\def\=#1{\underline{\underline #1}}

\def\E{{\bf E}}
\def\e{{\bf e}}

\def\h{{\bf h}}
\def\Eperp{{\bf E}_\perp}
\def\hperp{{\bf h}_\perp}
\def\Epara{{E}_\parallel}
\def\hpara{{h}_\parallel}

\def\eps{\epsilon}
\def\epst{{\boldsymbol \epsilon}}
\def\epsproj{\left.{\hat{\epst}}\right|}
\def\epsav{{\bar \epsilon}}
\def\Deps{{\Delta \epsilon}}

\def\Eperp{{\bf E}_\perp}
\def\hperp{{\bf h}_\perp}

\usepackage{hyperref}
\hypersetup{
    colorlinks=true,
    linkcolor=blue,
    citecolor=blue,
    urlcolor=blue,
    breaklinks=true,
}

\title{Broadband Bragg Phenomenon in a Uniform Medium}

\author{
  Martin W. McCall and Stefanos Fr. Koufidis \\
  Blackett Laboratory, Department of Physics, \\
   Imperial College of Science, Technology and Medicine, \\
  Prince Consort Road, London SW7 2AZ, United Kingdom\\
  \texttt{m.mccall@imperial.ac.uk} \\
}

\begin{document}
\maketitle

\begin{abstract}
A new mechanism of Bragg reflection is identified, one that, remarkably, occurs in a uniform medium and relies on resonant tuning of the medium's parameters. Due to uniformity, reflection ensues over a broad wavelength range, much like a metal, but it is polarization dependent: one circular state is reflected, whilst the other is transmitted. Such a medium can thus provide a broadband, low-loss polarization divider/combiner. Assessment of the required parameters suggests that manufacturing is within range of current metamaterials technology. By externally regulating the medium's parameters, a highly efficient optical modulator is possible with potential applications across optics, optoelectronics, and photonics.
\end{abstract}

\keywords{Birefringence \and Broadband devices \and Circular Bragg phenomenon \and Metamaterials \and Negative refraction \and Optical activity \and Polarization selectivity}

\section{Introduction}
Bragg reflection is said to occur when successive reflections of monochromatic light from a spatially modulated dielectric add up coherently. Such photonic structures are extensively used in optics, finding application, for example, in distributed feedback lasers \cite{Kogelnik1972}, in wavelength division multiplexers (WDM) \cite{Zengerle1995}, or as dispersion cancellation filters in optical waveguides \cite{Ouellette1987}. Fabrication of Bragg gratings \cite{Hill1993, Kashyap2009} is often challenging owing to the necessity of matching the scale of the medium's variation to the wavelength of light, say on the order of microns or less. In this letter, we identify a fundamentally new mechanism to achieve Bragg-like reflection that, remarkably, is independent of matching the wavelength to a spatial period, and relies solely on resonant tuning of the medium's parameters. The optical response is similar to the well-studied circular Bragg phenomenon (CBP) in structurally chiral media \cite{McCall2009}, whereby circularly polarized incident light that is co-handed with the medium's helicity is reflected, whilst contra-handed light is transmitted. The CBP has been observed naturally in iridescent beetle shells \cite{Michelson1911}, as well as in nano-fabricated sculptured thin films \cite{Lakhtakia2005}. However, unlike the traditional CBP, in our identified mechanism there is no “Bragg wavelength” as such, and polarization selective response occurs over a broad wavelength band, limited only by material dispersion. Evaluating the parameters necessary to achieve such optical response, we confirm that practical realisations are indeed within reach of current metamaterials technology.

\section{Combining birefringence with optical activity}
\par For a birefringent and optically active reciprocal medium, the temporal frequency domain constitutive relations combine a dielectric tensor $\epst$ with the Drude-Born-Fedorov model \cite{Bohren2003} 
\begin{align}\label{initial constitutive relations}
{\bf D}=\epsilon_0\left(\epst\cdot {\bf E}+i\alpha\eta_0{\bf H}\right)  \ \ \text{and} \ \ {\bf B}=\mu_0\left[-i\left({\alpha}/{\eta_0}\right){\bf E}+\mu{\bf H}\right]\,,
\end{align}
where $\bf E$, $\bf B$ are the fundamental electromagnetic fields, $\bf D$, $\bf H$ are the excitation fields, $\epsilon_{0}$, $\mu_{0}$, and $\eta_0=\left(\mu_0/\epsilon_0\right)^{1/2}$ are the free-space permittivity, permeability, and impedance,  respectively, and $\mu$ is the relative permeability. The chirality parameter $\alpha$ measures the length, in wavelengths, that a linearly polarized wave propagates before its $\bf{E}$-vector rotates through $2\pi$. Defining the auxiliary fields ${\bf h}=\eta_0{\bf H}$, ${\bf b}=(\eta_0/\mu_0){\bf B}$, and ${\bf d}=\epsilon_0^{-1}{\bf D}$, the constitutive relations of Eq. \eqref{initial constitutive relations} simplify to
\begin{align}\label{constitutive relations}
{\bf d}&=\epst\cdot {\bf E} + i\alpha {\bf h}  \ \  \text{and} \ \  {\bf b}= -i\alpha {\bf E} + \mu {\bf h}\,.
\end{align}
This notation assures that all fields are similarly dimensioned, so that all the medium's parameters are relative. We may now express the fields as $\E = \Epara {\hat {\bf k}}+\Eperp$ and $\h=\hpara {\hat {\bf k}}+\hperp$, with ${\hat {\bf k}}$ a unit vector in the direction of the wavevector $\bf k$ being perpendicular to  $\left(\Eperp,\hperp\right)$, and partition $\epst$ so that
\begin{equation}\label{partitioneps}
\epst \cdot \E =
\left(\begin{array}{c}{\hat{\boldsymbol\epsilon}}_\perp\cdot\Eperp +{\boldsymbol \epsilon}_\parallel \Epara\\ {\boldsymbol \epsilon}_\parallel^T\cdot\Eperp +\epsilon_s \Epara\end{array}\right)\,,
\end{equation}
where ${\hat{\boldsymbol\epsilon}}_\perp:{\mathbb C}^2 \rightarrow {\mathbb C}^2$, ${\boldsymbol \epsilon}_\parallel \in {\mathbb C}^2$, and $\epsilon_s \in {\mathbb C}$. Then, for a monochromatic plane wave ${\bf E}e^{i\left({\bf k}\cdot{\bf r}-\omega t\right)}$, combining the projections of Maxwell's macroscopic source-free curl relations yields
\begin{equation}\label{Exact Equation of E}
k^2\Eperp - 2i\alpha k_0k(\times)\Eperp - \mu^2 \epsproj\cdot\Eperp + \alpha^2 k_0^2\Eperp =0\,,
\end{equation}
where $\epsproj = {\hat{\boldsymbol\epsilon}}_\perp - \left(\epsilon_s-{\alpha^2}/{\mu}\right)^{-1}{\boldsymbol \epsilon}_\parallel {\boldsymbol \epsilon}_\parallel^T$ and in Cartesian coordinates $\left(\times\right)$ is interpreted as $\left(\begin{array}{cc}0&-1\\1&0\end{array}\right)$. The free-space wavenumber is $k_{0}={\omega}/{c}$, where $c$ is the speed of light in vacuum, and the wavenumber in the medium is $k$. In a propagation coordinate system (${{\bf k}} = k {\hat{\bf z}}$) with transverse coordinates ($x,y$) chosen to diagonalize $\epsproj$, as $\epsproj=\text{diag}\left(\epsilon_{1}, \epsilon_{2}\right)$, the supported by the medium refractive indices $n={k}/{k_0}$, turn out to be 
\begin{equation}\label{refreactive indices}
\pm n^{(\pm)}\equiv \pm\left[\epsav \mu +\alpha^2\pm \left(\Deps^2\mu^2+4\alpha^2\epsav\mu\right)^{1/2}\right]^{1/2}\,,
\end{equation}
where $\epsav=(\epsilon_1+\epsilon_2)/2$ is the average transverse dielectric constant and $\Deps=(\epsilon_2-\epsilon_1)/2$ measures the birefringence. 

\section{Wavelength-independent Bragg zones}
\par
\begin{figure}[!t]
\centering
\includegraphics[width= 0.5\linewidth]{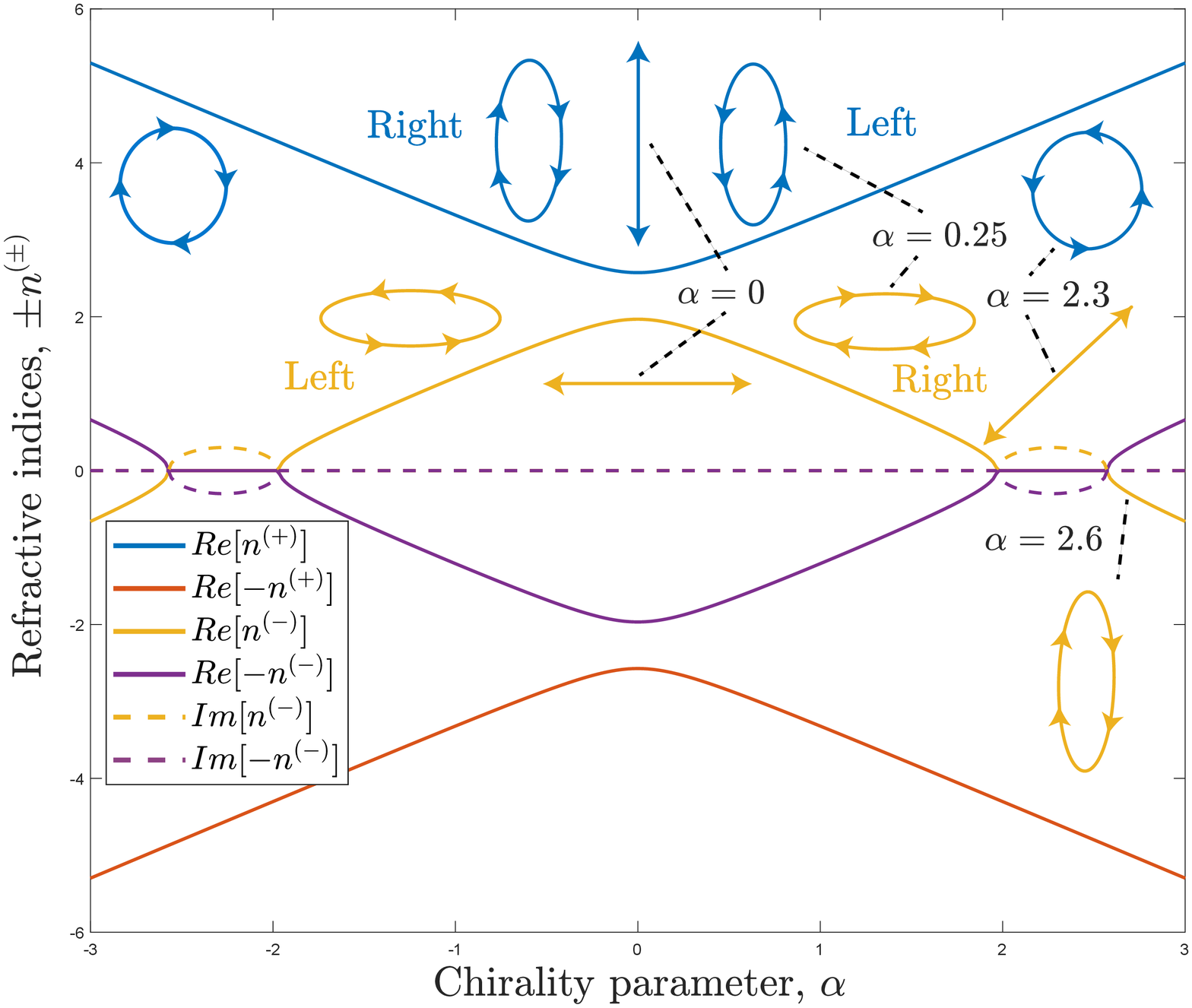}
\caption{
  Dispersion of a uniform birefringent and optically active medium in the chirality domain: the refractive indices of Eq. \eqref{refreactive indices} as functions of the chirality parameter, each associated with the polarization (linear, circular, or elliptical) of the corresponding eigenmode. Stopbands emerge around $\alpha\approx\pm\left(\bar{\epsilon}\mu\right)$. The base parameters are: $\epsilon_1=1.69$, $\epsilon_2=2.89$, and $\mu =\ 2.29$.}
\label{Figure Chirality Domain Dispersion}
\end{figure}

 For an idealized lossless medium with real parameters,   Eq. \eqref{refreactive indices} shows that $n$ is purely imaginary whenever 
\begin{equation}\label{stop band}
    \left(\epsav-\Delta\epsilon\right)\mu < \alpha^2 < \left(\epsav + \Delta\epsilon\right)\mu\,,
\end{equation}
assuming ordering such that $\Deps>0$. Within these bands, centered at $\alpha = \pm (\epsav\mu)^{1/2}$, one index is real and related to forward propagation, whilst the other is imaginary and related to evanescent waves. The sign of $\alpha$ determines which is which: for $\alpha>0$, as illustrated in Fig.\ \ref{Figure Chirality Domain Dispersion}, $n^{\left(+\right)}$ is associated with a left-handed propagating eigenmode, while $n^{\left(-\right)}$ is associated with a right-handed eigenmode which becomes evanescent in the stopband region $(\eps_1\mu)^{1/2}<\alpha<(\eps_2\mu)^{1/2}$. In this band, the eigenmode corresponding to $n^{\left(-\right)}$ becomes linear, while the eigenmode corresponding to $n^{\left(+\right)}$ remains circular. Accordingly, for $\alpha<0$, as also depicted in Fig. \ref{Figure Chirality Domain Dispersion}, the emerging stopband refers to a left-handed linear eigenmode.

\par Both stopband edges signify extreme values of chirality, with $\alpha$ being comparable to the average refractive index, $\left(\bar{\epsilon}\mu\right)^{1/2}$. This resonance condition triggers the onset of negative refraction in a purely optically active medium \cite{Zhang2009, Liu2021}. In fact, solving for optical activity, with a scalar $\epsilon$, the characteristic eigenvalues are $\pm\gamma_\pm=\pm k_0\left[\alpha\pm\left(\epsilon\mu\right)^{{1}/{2}}\right]$. For forward (backward) propagation, we must either choose $+\gamma_{+}$ or $-\gamma_{-}$ ($-\gamma_{+}$ or $+\gamma_{-}$). Then, the corresponding eigenmodes describe, nominally, right- and left- handed spatial helices. For $\left|\alpha\right|>\left(\bar{\epsilon}\mu\right)^{{1}/{2}}$, the direction of phase advance and handedness of two of the modes are interchanged, but the Poynting vector does not change \cite{McCall2009negativerefraction}.

\par
\begin{figure}[!t]
\centering
\includegraphics[width= 0.5\linewidth]{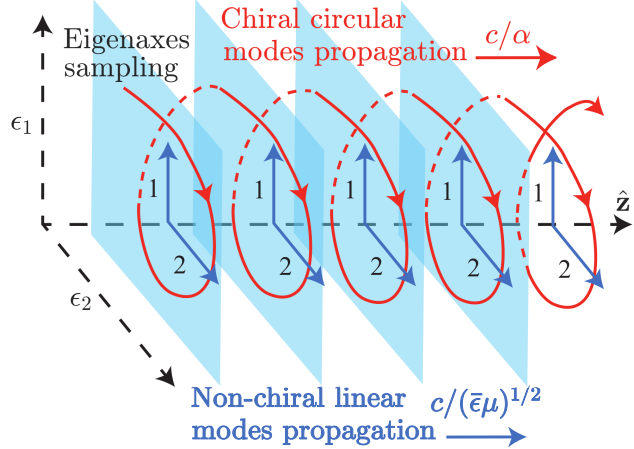}
\caption{  Heuristic picture: chiral circularly polarized eigenstates, propagating with speed $c/\alpha$, will synchronously and alternately sample the co-propagating eigenaxes if $\alpha \approx\left(\bar{\epsilon}\mu\right)^{1/2}$.}
\label{Figure Intuitive Picture}
\end{figure}

 An intuitive picture of how such a stopband occurs is illustrated in Fig.\ \ref{Figure Intuitive Picture}. In the absence of optical activity ($\alpha=0$), the polarization eigenstates are linear and propagate with speeds ${c}/{\left(\epsilon_{1,2}\mu\right)^{1/2}}$, or on average ${c}/{\left(\bar{\epsilon}\mu\right)^{1/2}}$. With optical activity present ($\alpha >0$, say), a circularly polarized wave propagating with speed $c/\alpha$ will synchronously and alternately sample the co-propagating eigenaxes, provided that $\alpha \approx\left(\bar{\epsilon}\mu\right)^{1/2}$. For $\Delta\epsilon\ll\bar{\epsilon}$, the purely imaginary refractive index of the resonant polarization is  $n_R^{\text{res}}\approx i\left({\Delta\epsilon}/{2}\right)\left({\mu}/{\bar{\epsilon}}\right)^{{1}/{2}}$, while the opposite-handed polarization, with refractive index $ n_{L}\approx\left(2\bar{\epsilon}\mu\right)^{1/2}$, propagates too slowly to sample the $\epsilon_{1,2}$ alternation synchronously. 

\section{Coupled wave theory description of birefringent and optically active media}
\par This interpretation is confirmed via coupled wave theory (CWT) \cite{Yariv1982}. To demonstrate this, we expand $\bf{E}$ in terms of circularly polarized states, $L$-left and $R$-right, for which the forward and backward propagating electric field amplitudes $A_{L,R}^\pm$ are perturbed along the direction of propagation $\hat{\bf{z}}$ by the presence of birefringence, i.e.,
\begin{equation}\label{Expansion of E}
      {\bf E}_\bot =\left(A_L^+e^{ik_0\left[\left(\bar{\epsilon}\mu\right)^{1/2}+\alpha\right]z}+A_R^-e^{-ik_0\left[\left(\bar{\epsilon}\mu\right)^{1/2}-\alpha\right]z}\right){\bf e}_{1} + \left(A_L^-e^{-ik_0\left[\left(\bar{\epsilon}\mu\right)^{1/2}+\alpha\right]z}+A_R^+e^{ik_0\left[\left(\bar{\epsilon}\mu\right)^{1/2}-\alpha\right]z}\right){\bf e}_{2}\,,
\end{equation}
where 
\begin{equation*}
   {\bf e}_{1}=\frac{1}{\sqrt2}\left(\begin{matrix}1\\i\\\end{matrix}\right) 
   \ \   \text{and} \ \
    {\bf e}_{2}=\frac{1}{\sqrt2}\left(\begin{matrix}1\\-i\\\end{matrix}\right)\,.
\end{equation*}
The relevant Helmholtz wave equation is found to be
\begin{equation}\label{Helmholtz Wave Equation}
    \frac{{\rm d}^2 {\bf E} }{{\rm d}z^2}+k_0^2\left(\mu{ {{{\epsproj}}}}-a^2\right)\cdot{\bf E}+2ak_0\hat{{\bf z}}\left(\times\right)\frac{{\rm d}{\bf E}}{{\rm d}z}={\bf 0}\,.
\end{equation}
Substituting Eq. \eqref{Expansion of E} into  Eq. \eqref{Helmholtz Wave Equation}, resolving along ${\bf e}_1$ and ${\bf e}_2$, and phase-matching potentially synchronous terms, we obtain two systems of coupled wave equations, one for each polarization state, namely
\begin{equation}\label{Coupled Wave Equations}
    \frac{\rm {d}}{{\rm d}z}\left(\begin{matrix}A_{L,R}^+\\A_{L,R}^-\\\end{matrix}\right)=i\kappa\left(\begin{matrix}0& e^{-ik_0\delta_{L,R}z}\\- e^{ik_0\delta_{L,R}z}&0\\\end{matrix}\right)\left(\begin{matrix}A_{L,R}^+\\A_{L,R}^-\\\end{matrix}\right)\,,
\end{equation}
with detuning parameters and coupling constant
\begin{equation}\label{CWE Parameters}
    \delta_{L,R}=2\left[\left(\bar{\epsilon}\mu\right)^{1/2}\pm\alpha\right]\ \   \text{and}\ \ \kappa\approx\frac{\pi}{\lambda_{0}}\left(\epsilon_1^{{1}/{2}}-\epsilon_2^{{1}/{2}}\right)\,,
\end{equation}
respectively, where $\lambda_0$ is the design wavelength. Crucially, on neglecting the dispersion of the average dielectric constant, the detuning parameters $\delta_{L,R}$ are \emph{wavelength-independent} being solely determined by the medium's parameters. The location of the resonances, each related to a specific polarization, can then be found by
setting $\delta_{L,R}=0$, leading to the regimes of Eq. \eqref{stop band}.

\par When $\alpha\approx\left(\bar{\epsilon}\mu\right)^{1/2}$ ($\alpha\approx-\left(\bar{\epsilon}\mu\right)^{1/2}$), the $A_L^\pm$ ($A_R^\pm$) amplitudes remain constant, while the $A_R^\pm$ ($A_L^\pm$) evolve exactly as for a Bragg grating (see, e.g., \cite{McCall2000}). Given a medium of length $z$, these evolutions are described by the solutions to Eq. \eqref{Coupled Wave Equations}
\begin{equation}\label{Coupled Wave Equations Solutions}
    \left(\begin{matrix}A_{L,R}^+\\A_{L,R}^-\\\end{matrix}\right)_{z}=\left(\begin{matrix}e^{- i\frac{k_0\delta_{L,R}z}{2}}P_{L,R}^+&e^{- i\frac{k_0\delta_{L,R}z}{2}}Q_{L,R}^+\\e^{ i\frac{k_0\delta_{L,R}z}{2}}Q_{L,R}^-&e^{ i\frac{k_0\delta_{L,R}z}{2}}P_{L,R}^-\\\end{matrix}\right)\left(\begin{matrix}A_{L,R}^+\\A_{L,R}^-\\\end{matrix}\right)_{0}\,,
\end{equation}
where 
\begin{align*}
   P_{L,R}^\pm&=\cosh{\left(q_{L,R}z\right)}\pm i\frac{k_0\delta_{L,R}}{2q_{L,R}}\sinh{\left(q_{L,R}z\right)}\,,\nonumber \\
    Q_{L,R}^\pm&=\frac{\pm i\kappa}{q_{L,R}}\sinh{\left(q_{L,R}z\right)\,,
     \ \  \text{and} \ \ q_{L,R}=\left[\kappa^2-\left(\frac{k_0\delta_{L,R}}{2}\right)^2\right]^{1/2}\,.
}
\end{align*}

\par
\begin{figure}[!t]
\centering
\includegraphics[width= 0.5\linewidth]{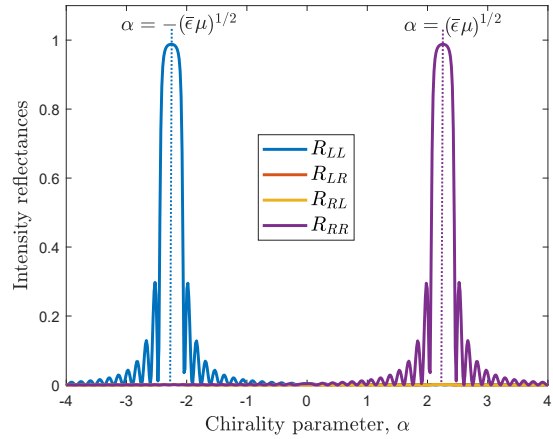}
\caption{  Intensity reflectances as functions of the chirality parameter for a thin film of a uniform birefringent and optically active medium. Light is normally incident and the surrounding medium is vacuum. The parameters are those of Fig.\ \ref{Figure Chirality Domain Dispersion} and the normalized film thickness is $k_{0}z=20$.}
\label{Figure Reflectances Spectrum}
\end{figure}

\begin{figure}[!t]
\centering
\includegraphics[width= 0.5\linewidth]{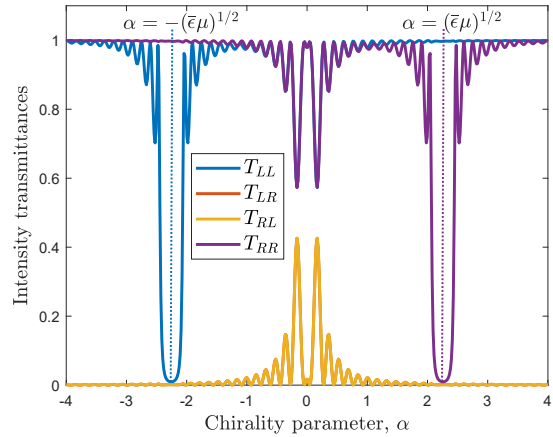}
\caption{  Intensity transmittances as functions of the chirality parameter for the scenario of Fig.\ \ref{Figure Reflectances Spectrum}. Around $\alpha\approx0$, an exchange between circular states ensues and at $\alpha=0$, the medium is linearly birefringent with purely linear eigenmodes.}
\label{Figure Tranmsittances Spectrum}
\end{figure}

 The optical response of a thin film of the considered medium to a normally incident plane wave may  then be computed analytically (via the eigenmodes of   Eq. \eqref{Exact Equation of E} and field-matching at both interfaces of the film, noting the subtlety that for $\alpha>\left(\epsilon\mu\right)^{{1}/{2}}$ ($\alpha <-\left(\epsilon\mu\right)^{{1}/{2}}$), it is the eigenmode associated with $n_{R}^{-}$ ($n_{L}^{+}$) that carries energy away from the interface) or approximately (via CWT, as in \cite{McCall2009}). The intensity reflectances $R_{ij}$ and transmittances $T_{ij}$, where $\{i,j\}=\{L,R\}$ indicate reflection or transmission of the $i$ polarization for incident $j$ polarization, are demonstrated in Figs.\ \ref{Figure Reflectances Spectrum} and \ref{Figure Tranmsittances Spectrum}, respectively. Bragg-like spectral features are evident at $\alpha\approx\pm\left(\bar{\epsilon}\mu\right)^{1/2}$, whereby for broadband incident light of equal amounts of right and left circular polarizations (RCP and LCP), half is strongly reflected for one polarization, while the other half is transmitted for the orthogonal one. Therefore, the medium acts as a broadband polarization divider. For $\alpha\approx\left(\bar{\epsilon}\mu\right)^{1/2}$ ($\alpha\approx-\left(\bar{\epsilon}\mu\right)^{1/2}$), the range over which the fraction of reflected intensity into RCP (LCP) for incident RCP (LCP), $R_{RR}$ ($R_{LL}$), is high ($>0.5$, say) is $\Delta\alpha=\left|\epsilon_2-\epsilon_1\right|\mu$. Within this range, unit reflectance of incident RCP (LCP) results in the amplitudes of forward and backward RCP (LCP) light being equal within the medium, the superposition of these giving a net linear polarization. Indeed, for $\alpha=(\bar{\epsilon}\mu)^{{1}/{2}}$ and just RCP light, combining Eq. \eqref{Expansion of E} with Eq. \eqref{Coupled Wave Equations Solutions} for $A_{R}^{-}\left(z\right)=0$, yields
\begin{equation}
    \frac{{\bf E}_\bot\left(0\right)}{A_R^+\left(0\right)}=\frac{i}{\sqrt2}\left(\begin{matrix}1\\i\\\end{matrix}\right)+\frac{1}{\sqrt2}\left(\begin{matrix}1\\-i\\\end{matrix}\right)=\frac{1+i}{\sqrt2}\left(\begin{matrix}1\\-1\\\end{matrix}\right)\,.
\end{equation}
Hence, the polarization is linear, which is logical as once $\alpha$ exceeds $(\epsilon_1\mu)^{1/2}$, the wave is evanescent and must match the backward propagating reflected field. 

\par
\begin{figure}[!t]
\centering
\includegraphics[width= 0.5\linewidth]{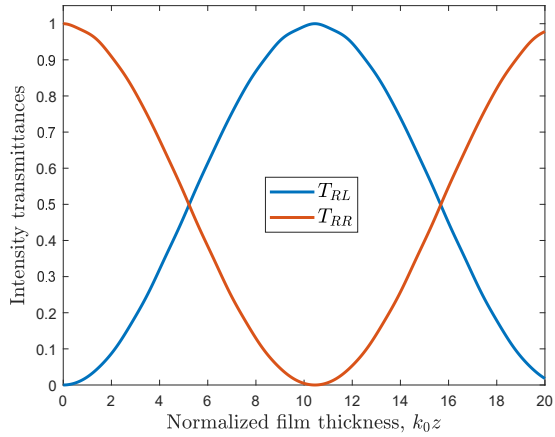}
\caption{  Continuous exchange between the $T_{RL}$ and $T_{RR}$ transmittances as a function of the normalized film thickness. The medium is uniform and birefringent with the parameters of Fig.\ \ref{Figure Chirality Domain Dispersion}, but with optical activity absent (cf. Fig. \ref{Figure Tranmsittances Spectrum}, $\alpha\approx0$ region).}
\label{Figure Exchange of Circular States}
\end{figure}

\begin{figure}[!t]
 \centering
\includegraphics[width=0.5\linewidth]{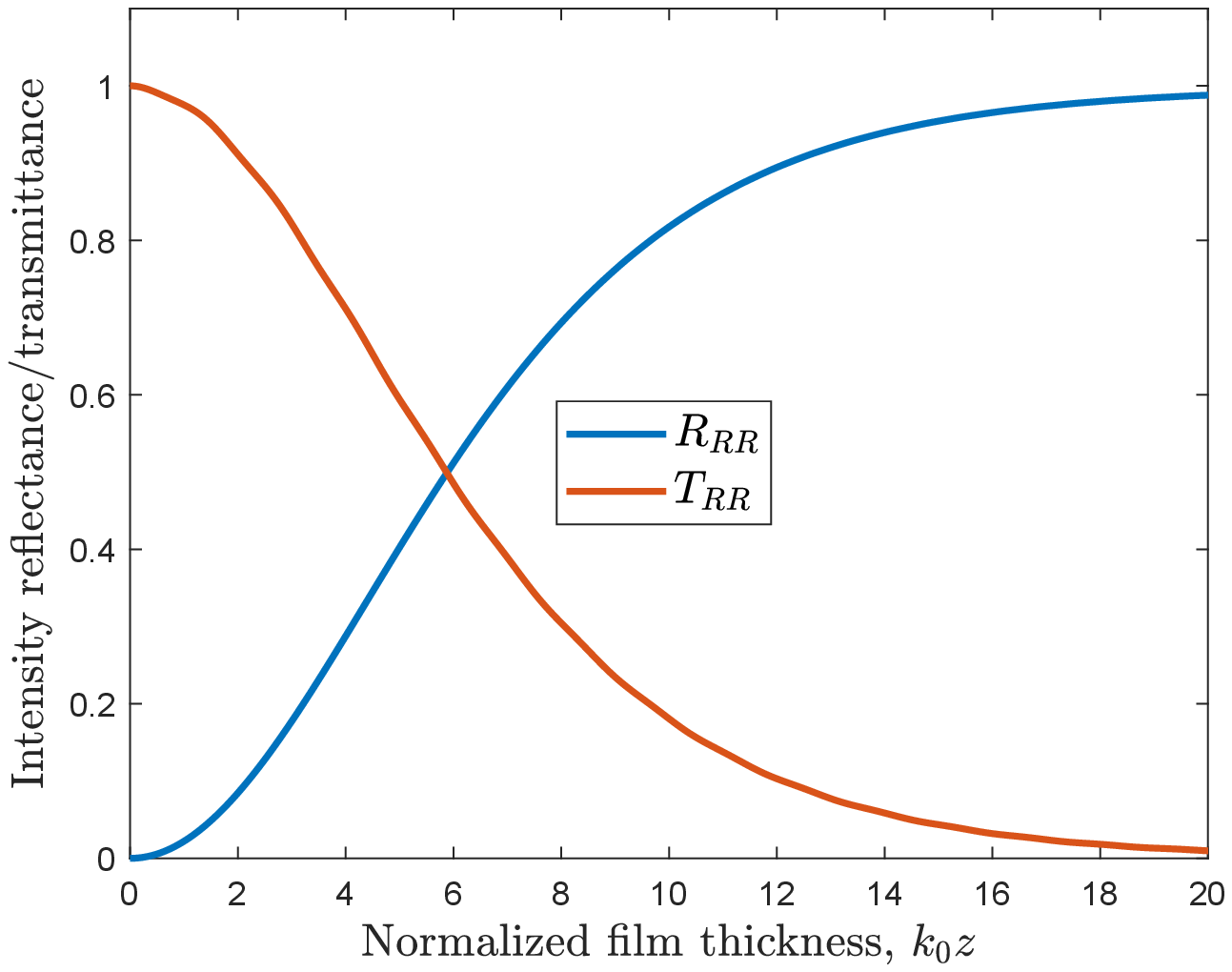}
\caption{  Bragg-like evolution of the $T_{RR}$ transmittance into the $R_{RR}$ reflectance as a function of film thickness in a uniform birefringent and optically active medium when on-resonance, at $\alpha=\left(\bar{\epsilon}\mu\right)^{1/2}$. The parameters are those of Fig.\ \ref{Figure Chirality Domain Dispersion}.}
\label{Figure Bragg-like Evolution of R and T}
\end{figure}

 Interestingly, for $\alpha\approx0$, resolving along ${\bf e}_{1}$ and ${\bf e}_{2}$, yet another pair of coupled wave equations emerges for the contra-handed forward (or backward) propagating waves. In terms of the circular components of the electric fields, and precisely at $\alpha=0$, both the exact and the CWT solutions agree to
\begin{equation}\label{Echange of Circular States}
    \left(\begin{matrix}E_L^+\\E_R^+\\\end{matrix}\right)_{z}=e^{ik_0(\bar{\epsilon}\mu)^{1/2}z}\left(\begin{matrix} \cos{\left(\kappa z\right)}&i\sin{\left(\kappa z\right)}\\i\sin{\left(\kappa z\right)}&\cos{\left(\kappa z\right)}\\\end{matrix}\right)\left(\begin{matrix}E_L^+\\E_R^+\\\end{matrix}\right)_{0}\,.
\end{equation}
This is a manifestation of the evolution of circularly polarized light in a linearly birefringent medium whereby $T_{RR}$ and $T_{LL}$ evolve, respectively, into $T_{RL}$ and $T_{LR}$, being `unnaturally' expanded on a circular basis. This continuous exchange of circular states between these transmissions with increasing film thickness is illustrated in Fig.\ \ref{Figure Exchange of Circular States}. By contrast, Fig.\ \ref{Figure Bragg-like Evolution of R and T} shows the characteristic Bragg-like evolution of $T_{RR}$ into $R_{RR}$ with increasing slab thickness, at $\alpha=\left(\bar{\epsilon}\mu\right)^{1/2}$, corroborating   Eq. \eqref{Coupled Wave Equations Solutions}.

\section{Material dispersion and practical realizations}
The foundations of the proposed Bragg-like phenomenon lie in the stopbands occurring in the medium's chirality domain dispersion, rather than that of wavelength. A natural question to address is whether considering material dispersion will affect the main features. Filling the medium under discussion with metallic spherical Mie resonators, uniformly distributed in space, as in \cite{Pendry2004}, and assuming the relative permittivities to be equal and dispersive in both birefringent axes, the stopbands appear, shifted, as long as the electron density  $N$,
\begin{equation}\label{electron density}
    N\ll N_{\text{max}}= \left(\epsilon_{0}m_{e}\right)^{{1}/{2}}q_{e}^{-1}\left|\omega^2-\omega_0^2\right|^{{1}/{2}}\,,
\end{equation}
where $m_e$, $q_e$ are the electron mass and charge, respectively, $\omega_{0}$ refers to the design frequency, and $\omega$ to the operational half-bandwidth. If, for example, the design wavelength is $\lambda_0=5$ \textmu m, and the desired bandwidth $\Delta\lambda=4$ \textmu m, then by setting $N=N_{\text{max}}\cdot{10}^{-3}$, the stopbands will be centred at $\alpha=\pm2.0602$ with $\Delta\alpha=0.7362$. Evidently, controlling $\omega_p$ via $N$, provides a \emph{switching mechanism} between resonant polarizations.

\par Although natural media that are simultaneously birefringent and optically active are known \cite{Moxon1990, Ghosh2008}, one with such extreme optical activity to satisfy $\alpha=\left(\bar{\epsilon}\mu\right)^{1/2}$ is unlikely to be found in Nature. Turning to metamaterials, the giant optical rotatory power of $\approx{10}^4$ $^\circ /{\rm mm}$, observed in \cite{Kuwata2005}, corresponds to a required refractive index of just $\left(\bar{\epsilon}\mu\right)^{1/2}=0.04$ at $\lambda_0=1.5$ \textmu m. Such extreme values have been observed in metamaterials at $\text{THz}$ frequencies ($\lambda_0 \approx 0.46$ \text{mm}) with the remarkably high index of $\left(\bar{\epsilon}\mu\right)^{1/2}=7$, achieving an optical rotatory power of $5.46\cdot{10}^3$ $^\circ /{\rm mm}$ in \cite{Liu2021}. Recently, embedding a chiral medium into a non-chiral gold nanorod metamaterial, enhanced the circular dichroism \cite{Zayats2018}. Hence, the phenomena predicted here are within reach of a carefully designed meta-medium. To reduce the required optical activity we may lower the average refractive index. Actually, one of the founding meta-media consisted of orthogonal wire inclusions that present a lower effective refractive index through control of the plasma frequency \cite{Rotman1962, Pendry1996}. Utilizing wires of different diameters in the orthogonal directions will also provide the necessary birefringence.

\section{Conclusion}
Concluding, we have demonstrated that a uniform birefringent and optically active medium exhibits a broadband and polarization selective Bragg phenomenon, exclusively relying on matching the medium's parameters rather than on wavelength matching. Current metamaterials technology provides the necessary parameters that are, moreover, in principle modulatable, leading to a very flexible broadband switch.  Potential applications include: lasers (broadband, short-pulsed), optoelectronics (OLED displays, electron transport in organic semiconductors),  waveguiding (add-drop WDM), or multi-use fiber sensors. 

\section*{Acknowledgments}
The authors acknowledge fruitful discussions with Dr. M. Foreman and Dr. K. Weir of Imperial College London.

\bibliographystyle{unsrt}  
\bibliography{references}

\end{document}